%-------------------------
% This paper uses harvmac
%-------------------------
\input harvmac.tex

%\def\subsubsec#1{$\underline{\rm #1}$}

% Something to deal with sub-sub-sections

\def\unlockat{\catcode`\@=11}
\def\lockat{\catcode`\@=12}

\unlockat
% Something to deal with sub-sub-sections

\def\newsec#1{\global\advance\secno by1\message{(\the\secno. #1)}
\global\subsecno=0\global\subsubsecno=0\eqnres@t\noindent
{\bf\the\secno. #1}
\writetoca{{\secsym} {#1}}\par\nobreak\medskip\nobreak}
\global\newcount\subsecno \global\subsecno=0
\def\subsec#1{\global\advance\subsecno
by1\message{(\secsym\the\subsecno. #1)}
\ifnum\lastpenalty>9000\else\bigbreak\fi\global\subsubsecno=0
\noindent{\it\secsym\the\subsecno. #1}
\writetoca{\string\quad {\secsym\the\subsecno.} {#1}}
\par\nobreak\medskip\nobreak}
\global\newcount\subsubsecno \global\subsubsecno=0
\def\subsubsec#1{\global\advance\subsubsecno by1
\message{(\secsym\the\subsecno.\the\subsubsecno. #1)}
\ifnum\lastpenalty>9000\else\bigbreak\fi
\noindent\quad{\secsym\the\subsecno.\the\subsubsecno.}{#1}
\writetoca{\string\qquad{\secsym\the\subsecno.\the\subsubsecno.}{#1}}
\par\nobreak\medskip\nobreak}

\def\subsubseclab#1{\DefWarn#1\xdef
#1{\noexpand\hyperref{}{subsubsection}%
{\secsym\the\subsecno.\the\subsubsecno}%
{\secsym\the\subsecno.\the\subsubsecno}}%
\writedef{#1\leftbracket#1}\wrlabeL{#1=#1}}% Macros for boxes
\lockat

\def\IL{\relax{\rm I\kern-.18em L}}
\def\IH{\relax{\rm I\kern-.18em H}}
\def\IR{\relax{\rm I\kern-.18em R}}
\def\IC{\relax\hbox{$\inbar\kern-.3em{\rm C}$}}
\def\IZ{\relax\ifmmode\mathchoice
{\hbox{\cmss Z\kern-.4em Z}}{\hbox{\cmss Z\kern-.4em Z}}
{\lower.9pt\hbox{\cmsss Z\kern-.4em Z}}
{\lower1.2pt\hbox{\cmsss Z\kern-.4em Z}}\else{\cmss Z\kern-.4em
Z}\fi}

\def\CB {{\cal B}}

%% MORE MACROS

\def\ch{{\rm ch}}

\font\manual=manfnt \def\dbend{\lower3.5pt\hbox{\manual\char127}}

\def\IZ{\relax\ifmmode\mathchoice
{\hbox{\cmss Z\kern-.4em Z}}{\hbox{\cmss Z\kern-.4em Z}}
{\lower.9pt\hbox{\cmsss Z\kern-.4em Z}}
{\lower1.2pt\hbox{\cmsss Z\kern-.4em Z}}\else{\cmss Z\kern-.4em
Z}\fi}
\def\half {{1\over 2}}

\def\ch{{\rm ch}}

% more macros, alphabetically

\def\IZ{\relax\ifmmode\mathchoice
{\hbox{\cmss Z\kern-.4em Z}}{\hbox{\cmss Z\kern-.4em Z}}
{\lower.9pt\hbox{\cmsss Z\kern-.4em Z}}
{\lower1.2pt\hbox{\cmsss Z\kern-.4em Z}}\else{\cmss Z\kern-.4em
Z}\fi}
\def\IB{\relax{\rm I\kern-.18em B}}
\def\IC{{\relax\hbox{$\inbar\kern-.3em{\rm C}$}}}
\def\ID{\relax{\rm I\kern-.18em D}}
\def\IE{\relax{\rm I\kern-.18em E}}
\def\IF{\relax{\rm I\kern-.18em F}}
\def\IG{\relax\hbox{$\inbar\kern-.3em{\rm G}$}}
\def\IGa{\relax\hbox{${\rm I}\kern-.18em\Gamma$}}
\def\IH{\relax{\rm I\kern-.18em H}}
\def\II{\relax{\rm I\kern-.18em I}}
\def\IK{\relax{\rm I\kern-.18em K}}
\def\IP{\relax{\rm I\kern-.18em P}}

\def\inbar{\,\vrule height1.5ex width.4pt depth0pt}

\def\mod{{\rm mod}}

\font\cmss=cmss10 \font\cmsss=cmss10 at 7pt
\def\IR{\relax{\rm I\kern-.18em R}}

\def\Tr{\rm Tr}

% Macros for boxes

\def\boxit#1{\vbox{\hrule\hbox{\vrule\kern8pt
\vbox{\hbox{\kern8pt}\hbox{\vbox{#1}}\hbox{\kern8pt}}
\kern8pt\vrule}\hrule}}
\def\mathboxit#1{\vbox{\hrule\hbox{\vrule\kern8pt\vbox{\kern8pt
\hbox{$\displaystyle #1$}\kern8pt}\kern8pt\vrule}\hrule}}

%% ANOTHER SET OF MACROS

\def\inbar{\,\vrule height1.5ex width.4pt depth0pt}

\font\cmss=cmss10 \font\cmsss=cmss10 at 7pt
\def\IR{\relax{\rm I\kern-.18em R}}

\def\Tr{\rm Tr}

%% NIKITAS MACROS

\def\a1{{\cal A}^{1,1}}

%REFERENCES
%
\lref\nov{ S. Novikov,"The Hamiltonian
formalism and many-valued analogue of  Morse theory",
Russian Math.Surveys 37:5(1982),1-56;
E. Witten, ``Global Aspects of Current Algebra,''
Nucl.Phys.B223:422,1983}
\lref\faddeevlmp{ L. D. Faddeev, ``Some Comments on Many Dimensional
Solitons'',
Lett. Math. Phys., 1 (1976) 289-293.}
\lref\polchinski{J. Polchinski, ``Dirichlet branes and Ramond-Ramond
charges,'' Phys. Rev. Lett. {\bf 75} (1995) 4724, hep-th/9510017.}
\lref\douglas{M. Douglas, ``Branes within branes,'' hep-th/9512077.}
\lref\strominger{A. Strominger, ``Open P-branes,'' hep-th/9512059.}
\lref\vw{C. Vafa and E. Witten, ``A One-Loop Test of
String Duality,'' Nucl. Phys. {\bf B447} (1995) 261, hep-th/9505053.}
\lref\polnotes{S. Chaudhuri, C. Johnson, and J. Polchinski,
``Notes on D-branes,'' hep-th/9602052.}
\lref\blumharva{J. Blum and J. A. Harvey, ``Anomaly inflow for gauge
defects,'' Nucl. Phys. {\bf B416} (1994) 119, hep-th/9310035.}
\lref\blumharvb{J. Blum and J. A. Harvey, unpublished.}
\lref\cnhvy{C. Callan and J. A. Harvey,  Nucl. Phys. {\bf B250}
(1985) 427. }
\lref\nacu{S. Naculich, Nucl. Phys. {\bf B296} (1988) 837.}
\lref\dlm{M. J. Duff, J. T. Liu and R. Minasian, ``Eleven dimensional
origin
of string/string duality: a one loop test,'' Nucl. Phys. {\bf B452}
(1995)
261, hep-th/9506126.}
\lref\librane{M. Li, ``Boundary states of D-branes and Dy-strings,''
Nucl. Phys. {\bf B460} (1996) 351, hep-th/9510161.}
\lref\bvs{M. Bershadsky, C. Vafa and V. Sadov, ``D-branes and
topological
field theories,'' hep-th/9511222.}
\lref\donaldson{S. Donaldson and P.B. Kronheimer,
{\it The Geometry of Four-Manifolds} Oxford 1990}
%\lref\polchinski{J. Polchinski, ``Dirichlet-Brances and
%Ramond-Ramond Charges,'' hep-th/9510017}
\lref\nov{ S. Novikov,"The Hamiltonian
formalism and many-valued analogue of  Morse theory",
Russian Math.Surveys 37:5(1982),1-56;
E. Witten, ``Global Aspects of Current Algebra,''
Nucl.Phys.B223:422,1983}
\lref\izt{J. M. Izquierdo and P. K. Townsend, Nuclear Physics {\bf B414}
(1994) 93.}
\lref\clny{C.G. Callan, C. Lovelace, C. R. Nappi and S. A. Yost, Nucl. Phys.
{\bf B308} (1988) 221.}
\lref\polcai{J. Polchinski and Y. Cai,
``Consistency of open superstring theories,''
Nucl.Phys. {\bf B296} (1988) 91.}
\lref\fadd{L. Faddeev, Phys. Lett. {\bf 145B} (1984) 81.}
\lref\dscnti{M.F. Atiyah and I.M. Singer,
``Dirac operators coupled to vector bundles,''
Proc. Natl. Acad. Sci. {\bf 81} (1984) 2597}
\lref\dscntii{
 L. Faddeev and S. Shatashvili, Theor. Math. Fiz., {\bf 60 }
(1984) 206; english transl. Theor. Math. Phys.
{\bf 60}(1984)770}
\lref\dscntiii{B. Zumino,
``Chiral anomalies and differential geometry,''
in {\it Relativity, Groups and Topology II},
proceedings of the
Les Houches summer school, Bryce S DeWitt and Raymond Stora, eds.
North-Holland, 1984.}
\lref\dscntiv{
For reviews see {\it Symposium on Anomalies, Geometry and Topology }
William A. Bardeen and Alan R.
White, eds. World Scientific, 1985, and
L. Alvarez-Gaum\'e and P. Ginsparg,
``The structure of gauge and gravitational anomalies,''
Ann. Phys. {\bf 161} (1985) 423.}
\lref\chs{C. Callan, J. A. Harvey and A. Strominger, ``Worldbrane actions
for string solitons,'' Nucl. Phys. {\bf B367} (1991) 60.}
\lref\ght{M. B. Green, C. M. Hull and P. K. Townsend, ``D-brane Wess-Zumino
actions, T duality and the cosmological constant,'' hep-th/9604119.}
\lref\br{E.  Bergshoeff and M.  de Roo, `D-branes and  T-duality',
hep-th/9603123.}

\Title{ \vbox{\baselineskip12pt\hbox{hep-th/9605033}
\hbox{DAMTP/96-40}
\hbox{EFI-96-13} \hbox{YCTP-P8-96}
\hbox{RU96-29}}}
{\vbox{
\centerline{I-Brane Inflow}
\centerline{and}
\centerline{Anomalous Couplings on D-Branes}
 }}
\medskip
\centerline{Michael Green $^1$, Jeffrey A. Harvey $^2$, and Gregory
Moore $^3$}

\vskip 0.5cm
\centerline{$^{1}$ DAMTP, Silver Street,  Cambridge CB3 9EW, UK }
\centerline{$^{2}$ Enrico Fermi Institute, University of Chicago}
\centerline{5640 Ellis Avenue, Chicago, IL 60637}
 \centerline{$^{3}$ Dept.\ of Physics, Yale University,
New Haven, CT  06520, Box 208120}
\vskip 0.1cm
\centerline{\bf Abstract}
We show that the anomalous couplings of $D$-brane gauge
and gravitational fields to Ramond-Ramond tensor potentials
can be deduced by a simple anomaly inflow argument applied
to intersecting $D$-branes and use this to determine the eight-form
gravitational coupling.

\Date{May 6, 1996}
%\draft

\newsec{Introduction}

Consider  a Weyl fermion on a $2n$-dimensional
manifold $\CB$ equipped with Yang-Mills
and gravitational connections $A, \omega$. If the fermions
transform in   a representation $\rho$ of
a gauge group the anomalous variation of
the action $\log Z(A, \omega)$  is given by the
famous descent formula \refs{\dscnti,\dscntii,\fadd,\dscntiii,\dscntiv}
 \eqn\descent{
\delta_\Lambda \log Z(A, \omega) =
2 \pi i \int_{\CB}
\biggl[ \ch_\rho(F) \hat A(R) \biggr]^{(1)}
}
where
\eqn\chdef{ \ch_\rho(F) = \Tr_\rho \exp({i F \over  2 \pi  } ) =
{\rm dim}~ \rho  +
\ch_1(F) + \ch_2(F) + \cdots }
($\ch_j$ is a $2j$-form)
and
\eqn\aroof{
\hat A(R) = 1 - { p_1\over  24} + {7 p_1(R)^2 - 4 p_2(R)\over  5760}
+ \cdots
 = 1 + \hat A_4(R) + \hat A_8(R) + \cdots
}
is the A-roof genus
\foot{Our conventions for curvatures and connections follows
\donaldson. Thus, $\ch_2$ is {\it negative} for an ASD
connection on a Euclidean 4-fold.}. In \descent\ we have used the
standard notation where for any closed gauge invariant form
$I$ we have $I - I_0 = d I^{(0)}$, where
$I_0$ is the leading constant term,
 and the first order gauge variation
is given by
$\delta I^{(0)} = d I^{(1)}$.  The descent
procedure is ambigous, reflecting the ability to add local
counterterms to the action. We will comment later on this
ambiguity in our context.

The formula \descent\ was given the following
simple physical
interpretation in \cnhvy.  Consider a $p$-brane  soliton
in a gauge/gravity theory in
an $s+1$-dimensional spacetime $X$. If $p$ is
odd it  can happen
that the soliton carries chiral fermions transforming
in an anomalous representation of the gauge/gravity theory
on the soliton with an anomaly determined by \descent\ with
$\CB$ the $p$-brane world volume.
In this case the effective theory on the soliton
violates charge conservation and (if $p=1 ~ \mod ~ 4$)
energy-momentum conservation. The apparent
charge violation - in a selfconsistent theory -
is accounted for by an inflow of charge
from the external nonanomalous theory. Equivalently, the gauge variation
of the low-energy effective action on the $p$-brane is cancelled
by a bulk term whose gauge variation is localized on the $p$-brane.

In this situation the bulk
theory has a $(p+1)$-form coupling to
$\CB$, i.e., there is a source term in the equations of
motion:
\eqn\object{
d * H_{p+2} = d\tilde H_{s-p-1}  = \delta_{s-p}(\CB  \rightarrow X)
\quad .
}
Here $\delta_{s-p}(\CB  \rightarrow X)$ is a delta-function supported Poincare
dual form of degree $s-p$.
The nonanomalous theory on $X$ has a
corresponding anomalous coupling
\foot{In the literature such terms are sometimes
called Chern-Simons couplings and sometimes
called WZ terms. We will simply adopt the name
``anomalous couplings.''}:
\eqn\bulk{
I^{\rm anom}_{\rm bulk} =
\int_{X  } \tilde H \wedge \biggl[ \ch_\rho(F) \hat
A(R)\biggr]^{(0)} .
}
We then see that the anomalous variation of $I^{\rm anom}_{\rm bulk}$
cancels the anomalous variation of the
effective action for the fermions on $\CB$ after using  the
equation of motion (Bianchi identity) \object
 \eqn\inflw{
\delta_\Lambda I^{\rm anom}_{\rm bulk} =
 \int_{\CB} \biggl[ \ch_\rho(F) \hat A(R)\biggr]^{(1) }.
}

Note that
this construction requires a correlation between the charge carried
by the $p$-brane and the number of fermion zero modes. Put another way,
given the fermion zero mode structure one can use this argument to
determine the quantum of charge carried by the $p$-brane.

This simple argument can be used to deduce the presence of chiral
fermion zero modes on the worldvolume given the presence of
the correct anomalous couplings, or it can be used, as we
do here,  to deduce the presence of anomalous couplings given
a knowledge of the fermion zero mode structure.

This construction also gives a satisfying picture of the role
played by consistent and covariant anomalies \nacu. The extension
of this mechanism to theories with Green-Schwarz anomaly cancellation
was discused in \blumharva\ and the relation between this mechanism of anomaly
cancellation and the Green-Schwarz mechanism was described in \izt.
These couplings have  played a crucial role in the study of
certain string dualities.
For example both Type IIA string theory and $M$ theory  have
fivebranes with a chiral and anomalous spectrum \chs.
Cancelling this anomaly on the fivebrane by a bulk inflow
determines a coupling between the potential coupling to the fivebrane
and an 8-form polynomial in curvature {\blumharvb, \dlm}
as can be verified in the IIA theory by a direct string calculation \vw.

In this note we show how the same line of argument
determines the anomalous couplings of
gauge brane fields with bulk fields and verifies Polchinski's
calculation of
the quantum of Ramond-Ramond charge carried by $D$-branes \polchinski.
For $N$
coincident D p-branes on $\CB_p$ we find the anomalous couplings
\eqn\branecs{I^{CS}
 = \int_{\CB_p }  C \wedge  \Tr_{N} \exp({ i F \over  2 \pi  })
\sqrt{\hat A(R)}}
with
\eqn\iidefs{\sqrt{\hat A(R)} = 1 + \half \hat A_4 + (\half \hat A_8 -
{1 \over 8} {\hat A}_4^2 ) + \cdots }
where $F$ is the $U(N)$
gauge field strength localized on the brane and $R$ is the
10-dimensional Riemann curvature 2-form pulled back to the
brane worldvolume and
the RR  field strength $H=dC+ \cdots $ is a
bispinor, equivalent to a sum of even (odd) forms for
IIA (IIB) string theory.

The gauge field terms in \branecs\ were first found in
\librane\ using the results of \refs{\polcai, \clny} and have
a number of applications to string duality \refs{\douglas,
\strominger}.
The $\hat A_4$ term was deduced in \bvs\ using a ``duality
chasing'' argument and plays a crucial role in IIA-heterotic
duality. We will verify these terms below.
The result for the degree eight gravitational coupling
is new. In \branecs\ the
Neveu-Schwarz potential, $B$, has been
set to zero for  convenience although non-zero
$B$ can easily be included in the following.

\newsec{Derivation}

Consider two orthogonally intersecting type II Dirichlet $p$-branes
of dimensions $p_1, p_2$, multiplicities
$N_1, N_2$, and filling worldvolumes $\CB_1, \CB_2$.   We will assume
they lie along coordinate subspaces and intersect in an I $p_{12}$-brane
$\CB_{12} = \CB_1 \cap \CB_2 $ (We use the nomenclature I-brane to indicate
the brane occuring at the interesection of two D-branes. The I-brane
zero modes differ from those of a D-brane of the same dimension).
Accordingly, we may split up the spacetime coordinates into
mutually disjoint sets $\{ 0, \dots 9\} = S_{12}\amalg S_1\amalg S_2
\amalg T$.
String endpoints  in $\CB_1$ have Neumann (N) boundary
conditions in $S_{12} \amalg S_1$ and Dirichlet (D) boundary
conditions in $S_{2} \amalg T$ while enpoints in $\CB_2$
have Neumann boundary conditions in $S_{12} \amalg S_2$ etc.
The I-brane   $\CB_{12}$  lies
along the coordinate plane defined by $S_{12}$.

We are interested in the case when the I-brane possesses chiral,
anomalous zero modes. It is not hard to see that this happens when
there are chiral unbroken supersymmetries on the I-brane.
Unbroken supersymmetries
will exist if  there exist $SO(1,9)$ MW
spinors $\epsilon, \tilde \epsilon$ such that \polnotes:
\eqn\mwsp{
\eqalign{
\epsilon & = \Gamma^{S_{12} } \Gamma^{S_1} \tilde \epsilon\cr
\epsilon & = \pm   \Gamma^{S_{12} } \Gamma^{S_2} \tilde \epsilon\cr}
}
Each
equation in \mwsp\ has solutions iff  $p_1,p_2$ are both
even (odd) in the IIA (IIB) theory. Moreover, each linearly
independent solution of $\epsilon = \eta \Gamma^{12} \Gamma^T
\epsilon$
(where $\eta$ is a sign depending on $S_{12},S_1,S_2$) gives a
linearly independent supersymmetry. Solutions exist iff
  $\vert S_1 \vert + \vert S_2 \vert = 0 ~\mod ~4$. This may be
proved by squaring the $\Gamma$-matrix or by noting that
$\vert S_1 \vert + \vert S_2 \vert  $ is invariant under
$T$-duality  transformations along all coordinate axes. Therefore
we may map one configuration to a D-instanton and apply
the result of \polnotes.  The supersymmetry can only be chiral for
$p_{12}=1~\mod~ 4$ and $T= \emptyset$. Up to $T$-duality
there are exactly two distinct cases with chiral supersymmetry.
We can have two
5-branes intersecting on a string:
$S_{12}=\{ 0,1\}, S_1 = \{ 2,3,4,5\}, S_2= \{6,7,8,9\}$
or we can have two 7-branes intersecting on a
5-brane:
$S_{12}=\{ 0,1,2,3,4,5 \}, S_1 = \{ 6,7\}, S_2= \{ 8,9\}$.

The excitation spectrum of the brane theory is easily derived
using the techniques explained in  \polnotes. There
are four sectors with boundary conditions:
\eqn\mxdsect{
\eqalign{
\matrix{   & 	S_{12} & 	 S_1 & S_2  & T \cr
     	\sigma=0 	& N & 	N & D   & D \cr
     	\sigma=\pi 	& N & 	 N& D   & D \cr}
\quad
&
\quad
\matrix{   & 	S_{12} & 	 S_1 & S_2  & T \cr
     	\sigma=0 	& N & 	D & N   & D \cr
     	\sigma=\pi 	& N & 	D & N   & D \cr}
\cr
& \cr
\matrix{   & 	S_{12} & 	 S_1 & S_2  & T \cr
     	\sigma=0 	& N & 	D & N   & D \cr
     	\sigma=\pi 	& N & 	N & D   & D \cr}
\quad
&
\matrix{   & 	S_{12} & 	 S_1 & S_2  & T \cr
     	\sigma=0 	& N & 	N & D   & D \cr
     	\sigma=\pi 	& N & 	D & N   & D \cr}
\quad
\cr}
}
The first two sectors  lead to
$U(N_i)$ SYM on $\CB_i$.
The  second two sectors provide supermultiplets
in the $(N_1, \bar N_2)$ and $(\bar N_1,   N_2)$
of the gauge group $U(N_1) \times U(N_2)$.
These fields, corresponding to the two orientations
of DN strings, give fields related by complex
conjugation.
Evidentally, these ``mixed sector''
fields only have zeromodes along
$\CB_{12}$.  There are always massless states in the Ramond
sector and, by the GSO projection, they will be chiral fermions
if and only if $p_{12}$ is odd (regardless of whether we work in
IIA or IIB theory).  Since the states confined to $\CB_{12}$
only come from open strings we never encounter
chiral bosons or gravitini.  In short, the massless
spectrum  on the I-brane consists of one Weyl fermion in
the $(N_1, \bar N_2)$ and one in the $(\bar N_1, N_2)$.

In four or eight world-volume dimensions this spectrum is not
chiral since complex conjugation flips the chirality and there is
thus no anomaly on the I-brane.
In two or six world-volume dimensions
there is an anomaly determined by descent from the four-form
or eight-form part of
\eqn\defrom{I = (\ch_{(\bar N_1, N_2)} (F) + \ch_{(N_1, \bar N_2)}(F) )
\hat A(R)
\cong
2 \ch_{N_1}(F_1) \ch_{N_2} (F_2) \hat A(R) }
where we have used the fact that in two or six dimensions only even powers
of the gauge field strength appear so that traces in the $N$ and $\bar N$
are equal.

Now
let us compare the anomalous charge
violation \defrom\ on the I-brane with the variation coming
from the bulk terms on the two intersecting branes. We will follow
the convention of \polchinski\ with two parameters $\mu$, $\alpha$
appearing in the normalization of the RR kinetic terms and the coupling
of the RR potential to D-branes, and we also take $4 \pi^2 \alpha' = 1$.
We provisionally assume a coupling
on each brane of the form
\eqn\branecoup{\mu \int_{\CB} C \wedge Y(F,R) }
where $Y(F,R)$ is a gauge invariant polynomial
of mixed degree with $Y(F,R) = N + \cdots$.
Strictly speaking, the coupling \branecoup\
is not well-defined
in the presence of branes
since the RR potentials $C$ are not
single valued or mutually local. This can be remedied by integrating
by parts all terms except for the top RR potential.
Thus \branecoup\ is more properly written as
\eqn\mrprpr{
\mu \int_{\CB} N C+ Y^{(0)} H  \quad . }
Since $H$ and
$dC$ differ in the presence of branes
($H$ is gauge invariant) \mrprpr\  in fact
differs from \branecoup. We consider
\mrprpr\ as the correct expression of the
brane anomalous coupling.

The equation of motion (Bianchi identity) for the
RR field strength is:
\foot{In order to avoid factors of $\sqrt{-1}$ we
work in Minkowski space for IIB and Euclidean
space for IIA.}
\eqn\rreq{dH = {\mu \over \alpha} \sum_{\rm branes}
\delta(\CB_p \rightarrow M_{10}) Y(F,R) .
}
The RR potential $C$ thus has an anomalous variation
in the presence of branes given by
\eqn\cvar{
\delta C = - {\mu \over  \alpha}
\sum_{\rm branes}
\delta(\CB_p \rightarrow M_{10}) Y^{(1)}(A,\omega)
}
where $A$ is the brane gauge field.

We can now compute the variation of \mrprpr\ for two intersecting
branes using \rreq\ and \cvar\ and find a variation on the I-brane of
\eqn\ibvar{
- {\mu^2 \over  \alpha} \int_{\CB_{12}}
\Biggl[ Y(1) Y^{(1)}(2) + Y(2) Y^{(1)}(1)
+ N_1 Y^{(1)}(2) + N_2 Y^{(1)}(1) \Biggr]
}
where $Y(i)$ indicates that it is a
function of $\omega$ and the gauge
fields on the $i^{th}$ brane.
The anomalous variation \ibvar\  follows
from descent from the polynomial
\eqn\ibfrom{ - {\mu^2 \over  2 \pi \alpha} 2 Y(1) Y(2) . }
Symmetry between the two branes fixes the
local counterterm ambiguity to give:
\eqn\defdesc{
(2 Y(1) Y(2))^{(0)}   = Y(1)^{(0)} Y(2) + Y(1) Y^{(0)}(2)
  + N_1 Y^{(0)}(2)  +N_2 Y^{(0)}(1).  }
Comparing this to the anomaly on the I-brane \defrom\ we see that the
anomaly is cancelled provided that
\eqn\answr{
\eqalign{
Y(F,R) & = \ch_N(F) \sqrt{\hat A(R)} , \cr
\mu^2/\alpha & = 2 \pi , \cr}
}
thus verifying \branecs\ as well
as the quantum of RR charge found in \polchinski.

Quantization of $\mu$  constrains, in particular, the nine-form
potential whose  source is the eight-brane, and  the Romans mass  of type IIA
ten-dimensional supergravity. However, when this mass is
non-zero there are  extra terms in \answr\ that should be easy to
incorporate following \br\ and \ght.   Likewise
it is easy to include a
non-zero Neveu--Schwarz
potential which was  set equal to zero in the above.

It is natural to wonder about the inflow mechanism
for multiply intersecting branes. However,
since open strings have only two
ends all charge violation is already accounted
for by considering pairs of branes.

\centerline{\bf Acknowledgements}

We would like to thank the Rutgers
department of physics for hospitality
during the course of this work.
In particular we would like to thank
T. Banks, M. Douglas, A. Losev,  E. Martinec,
N. Nekrasov, A. Sen, and S. Shatashvili
for useful discussions.
The research of GM
was supported by DOE grant DE-FG02-92ER40704,
and by a Presidential Young Investigator Award.
The work of JH was  supported in part by NSF Grant No.~PHY 91-23780.

\listrefs

\bye